\documentclass[fleqn,usenatbib]{mnras}

\usepackage{newtxtext,newtxmath}
\usepackage[T1]{fontenc}

\DeclareRobustCommand{\VAN}[3]{#2}
\let\VANthebibliography\thebibliography
\def\thebibliography{\DeclareRobustCommand{\VAN}[3]{##3}\VANthebibliography}

\usepackage{graphicx}
\usepackage{amsmath}

\title[Constraining baryonic feedback and cosmology]{Constraining baryonic feedback and cosmology with weak-lensing, X-ray, and kinematic Sunyaev-Zeldovich observations}

\author[Schneider et al.]{
Aurel Schneider,$^{1}$\thanks{E-mail: aurel.schneider@uzh.ch}
Sambit K. Giri,$^{1}$
Stefania Amodeo,$^{2}$ and Alexandre Refregier$^{3}$
\\
$^{1}$Institute for Computational Science, University of Zurich,
Winterthurerstrasse 190, 8057 Zurich, Switzerland\\
$^{2}$Universit\'{e} de Strasbourg, CNRS, Observatoire astronomique de Strasbourg, UMR 7550, F-67000 Strasbourg, France\\
$^{3}$Institute for Particle Physics and Astrophysics, ETH Zurich, Wolfgang Pauli Strasse 27, 8093 Zurich, Switzerland
}

\date{Accepted XXX. Received YYY; in original form ZZZ}

\pubyear{2021}

\begin{document}
\label{firstpage}
\pagerange{\pageref{firstpage}--\pageref{lastpage}}
\maketitle

\begin{abstract}
Modern weak-lensing observations are becoming increasingly sensitive to baryonic feedback processes which are still poorly understood. So far, this challenge has been faced either by imposing scale-cuts in the data or by modelling baryonic effects with simple, one-parameter models. In this paper, we rely on a more general, seven-parameter prescription of baryonic feedback effects, which is primarily motivated by observations and has been shown to agree with a plethora of hydrodynamical simulations. By combining weak-lensing data from the Kilo-Degree Survey (KiDS-1000) with observations of gas around galaxy clusters, we are able to constrain baryonic parameters and learn more about feedback and cosmology. In particular, we use cluster gas fractions from X-ray data and gas profiles from kinematic Sunyaev-Zeldovich (kSZ) observations to provide evidence for baryonic feedback that is stronger than predicted by most hydrodynamical simulations. In terms of the matter power spectrum, we report a beyond-percent effect at wave-modes above $k\sim 0.1-0.45$ h/Mpc and a maximum suppression of $12-33$ percent at $k\sim7$ h/Mpc (68 percent confidence level). Regarding the combined parameter $\Sigma_8=\sigma_8(\Omega_m/0.3)^{0.58}$, we find the known tension with the Planck satellite data to be reduced from 3.8 to 2.9 $\sigma$ once baryonic effects are fully included in the analysis pipeline. The tension is further decreased to 2.6 $\sigma$ when the weak-lensing data is combined with X-ray and kSZ observations. We conclude that, while baryonic feedback effects become more important in modern weak-lensing surveys, they are unlikely to act as the main culprit for the observed $\Sigma_8$-tension.
\end{abstract}


\section{Introduction}
The standard model of cosmology ($\Lambda$CDM) based on the cosmological constant ($\Lambda$) plus cold dark matter (CDM) has so far shown to be very successful in predicting cosmological observations from many different scales and redshifts. Despite this obvious success, the model remains unsatisfactory in the sense that the nature of its two main components is not truly understood. One of the main goals of current and upcoming cosmological surveys is to stress-test the $\Lambda$CDM model using different observational probes from various scales and redshifts. Any appearing discrepancy between different observations could point towards new physics and might ultimately lead to a better understanding of the dark sector. 

Over the last decade, low-redshift probes of the cosmological large-scale structure have substantially improved in accuracy. As a result, a mild yet persistent tension has appeared between the measured amplitudes of matter fluctuations compared to results from the cosmic microwave background (CMB). More specifically, the combined cosmological parameter $S_8=\sigma_8(\Omega_m/0.3)^{0.5}$ obtained by the weak-lensing surveys {\tt CFHTLens} \citep{Heymans:2013fya,Joudaki:2016mvz}, {\tt KiDS} \citep{Hildebrandt:2016iqg, KiDS:2020suj}, {\tt DES} \citep{DES:2017qwj,DES:2021bvc}, and {\tt HSC} \citep{HSC:2018mrq} is consistently low compared to the CMB value from {\tt Planck} \citep{Planck:2018vyg}. The tension is only at the 2-3.5 $\sigma$ level, and therefore statistically not very significant, but it appears consistently in all weak-lensing surveys. Other low-redshift observables, such as cluster counts \citep[][]{Planck:2015lwi,Corasaniti:2021ihg} and CMB-lensing \citep{Planck:2018lbu,SPT:2019fqo}, show similar trends, albeit at even lower statistical significance.

One potential systematic that low-redshift probes such as weak lensing, galaxy clustering, and cluster count measurements have to deal with is baryonic feedback, i.e. gas ejection driven by energy blasts from active galactic nuclei (AGN). Baryonic feedback leads to a redistribution of gas, altering the nonlinear density field at cosmological scales \cite[see e.g.][for a comprehensive review]{Chisari:2019tus}. Based on the results from hydro-dynamical simulations, a suppression of the matter power spectrum of order 10 percent at around $k\sim 1$ h/Mpc is expected \citep{vanDaalen:2011xb, Semboloni:2011aaa}. However, the exact amplitude of the effect remains unknown \citep{Schneider:2018pfw,vanDaalen:2019pst}.

Current weak lensing surveys have either followed the strategy of modelling baryonic feedback using an additional free parameter (as e.g. done by {\tt KiDS} and {\tt HSC}) or ignoring data from small scales, where feedback effects are believed to be significant an approach pursued in the standard {\tt DES} analysis papers). Both approaches rely on knowledge of the amplitude of feedback at different cosmological scales which are taken from hydro-dynamical simulations. However, simulations include AGN feedback as a subgrid model and cannot provide predictions that are independent of of modelling choices. The current results from weak-lensing therefore carry a certain amount of uncertainty that is difficult to quantify. 

In this paper we use an alternative approach to deal with the baryonic feedback problem. Instead of relying on results from hydro-dynamical simulations, we directly constrain baryonic effects using observed gas distributions around galaxy groups and clusters. We thereby use the baryonification model \citep{Schneider:2015wta,Schneider:2018pfw, Giri:2021qin} which provides a direct connection between halo properties (such as the gas, stellar, and dark matter profiles) and the matter power spectrum (and thus the weak-lensing signal). This allows us to perform a cosmological parameter inference study including weak-lensing data together with X-ray observations from galaxy clusters as well as the kinematic Sunyaev-Zeldovich (kSZ) signal. 

We argue this to be the first study to self-consistently include baryonic feedback effects into a cosmological inference analysis without relying on hydrodynamical simulations that could include hidden biases due the implemented AGN subgrid model. On the one hand, this approach allows us to confirm the viability of previous studies regarding baryonic feedback. On the other hand, we are able to investigate if direct gas observation provide evidence for particularly strong baryonic feedback that could alleviate the $S_8$-tension.

Next to the $S_8$-tension, we explore the parameter constraints of the baryonification model obtained from the {\tt KiDS} weak-lensing data alone and from combinations of weak lensing with X-ray observations and the %kinematic Sunyaev-Zeldovich 
kSZ effect. Furthermore, we derive the first constraints on the baryonic suppression of the matter power spectrum, which include uncertainties from both baryonic physics and cosmology. These model-independent constraints allow us to carry out a direct comparison to results from hydrodynamical simulations.

The paper is organised as follows: In Sec.~\ref{sec:BFCmodel} we summarise the baryonification model with a focus on its free parameters that will be varied during the inference analysis. Sec.~\ref{sec:data} provides an overview of the data products for the weak-lensing, the X-ray, and the kSZ measurements. We especially show how baryonic feedback processes affect these observations. In Sec.~\ref{sec:results} we present the results, showing in particular the effect of baryonic feedback on the $S_8$ tension and the matter power spectrum. Sec.~\ref{sec:conclusions} finally provides a conclusion of the paper as well as an outlook towards upcoming weak-lensing surveys.

\section{Baryonification model}\label{sec:BFCmodel}
The baryonification method relies on an approach to perturbatively shift simulation particles in gravity-only $N$-body outputs in order to mimic the presence of gas and stars. The goal of the particle shifts is to transform the original dark-matter-only profiles ($\rho_{\rm dmo}$) into more realistic dark-matter-baryon profiles ($\rho_{\rm dmb}$)
\begin{equation}
\rho_{\rm dmo}\,\,\,\rightarrow\,\,\,\rho_{\rm dmb}=\rho_{\rm clm}+\rho_{\rm gas}+\rho_{\rm cga},
\end{equation}
where the latter consists of a collision-less matter ($\rho_{\rm clm}$), a gas ($\rho_{\rm gas}$), and a central-galaxy component ($\rho_{\rm cga}$). The collision-less matter profile is dominated by dark matter but also contains a stellar part from satellite galaxies. The gas and stellar profiles consist of parametrised functions that are motivated by observations \citep{Schneider:2015wta}. They are furthermore back-reacting on the collision-less profile in a process referred to as adiabatic contraction and expansion \citep[see e.g.][]{Teyssier:2010dp}. 

We refer to \cite{Schneider:2018pfw} and \cite{Giri:2021qin} for a detailed description of the baryonification model. Here we only show some key quantities such as the gas profile
\begin{eqnarray}\label{rhogas}
\rho_{\rm gas}(r)\propto\frac{(\Omega_b/\Omega_m) -f_{\rm star}(M_{\rm vir})}{\left[1+10\left(\frac{r}{r_{\rm vir}}\right)\right]^{\beta(M_{\rm vir})}\left[1+\left(\frac{r}{\theta_{\rm ej}r_{\rm vir}}\right)\right]^{[\delta-\beta(M_{\rm vir})]/\gamma}}
\end{eqnarray}
with mass-dependent slope
\begin{equation}
\beta(M_{\rm vir}) = \frac{3\left(M_{\rm vir}/M_{c}\right)^{\mu}}{1+\left(M_{\rm vir}/M_{c}\right)^{\mu}} \ ,
\end{equation}
where $r_{\rm vir}$ and $M_{\rm vir}$ are the virial radius and mass, respectively.
Eq.~(\ref{rhogas}) describes a cored power-law with a steep truncation beyond $r_{\rm ej}=\theta_{\rm ej}r_{\rm vir}$. For the largest clusters ($M_{\rm vir}\gg M_c$), the slope of the power-law approaches $\beta=3$, which means that the profile follows a (truncated) NFW profile. This is not the case for smaller clusters and galaxy groups, where the slope becomes significantly shallower ($\beta<3$), mimicking the effects of baryonic feedback processes.

The stellar-to-halo fractions of the central galaxy ($f_{\rm cga}$) and the total stellar content ($f_{\rm star}$) are given by
\begin{equation}\label{fstar}
f_{i}(M_{\rm vir}) = \frac{M_{i}}{M_{\rm vir}} = 0.055\left(\frac{M_{s}}{M_{\rm vir}}\right)^{\eta_i}
\end{equation}
with $i=\lbrace {\rm cga},{\rm star}\rbrace$. Here we asssume $M_s=2\times10^{11}$ M$_{\odot}$/h as well as $\eta_{\rm star}\equiv \eta$ and $\eta_{\rm cga}\equiv \eta+\eta_{\delta}$. Note that Eq.~(\ref{fstar}) is motivated by the large-scale behaviour of the fitting function from \cite{Moster:2013aaa}.

The baryonification model summarised above comes with seven free model parameters, five describing the gas distribution ($M_c$, $\mu$, $\theta_{\rm ej}$, $\gamma$, $\delta$) and two the stellar abundances ($\eta$, $\eta_{\delta}$). Although it is shown in \citet{Giri:2021qin} that three parameters are sufficient to recover the results from known hydrodynamical simulations, we will simultaneously vary all seven parameters in this paper. This is a very conservative choice motivated by the fact that we do not know if current simulations properly model the gas distribution at cosmological scales.

One of the main advantages of the baryonification model is that it connects the distribution of gas and stars around haloes with statistics of the large-scale structure, such as the matter power spectrum. This allows us to not only quantify the effects of baryonic feedback on weak-lensing statistics, but to connect weak-lensing results with observations of gas around galaxy groups and clusters. We will thereby focus on observations from X-ray gas fractions and gas profiles measured by the kSZ effect.

Note that for this paper, whenever we require information about the matter power spectrum, we will use the baryonic emulator of \cite{Giri:2021qin} instead of directly applying the baryonification model to $N$-body simulations. The baryonic emulator has been built using 2700 baryonified simulation outputs, randomly distributed over the full parameter space. It provides percent accurate predictions of the baryonic power suppression signal and is fast enough to be readily used for a multi-parameter Bayesian analysis.

\begin{figure*}
\centering
\includegraphics[width=1.0\textwidth,trim=0.5cm 0.3cm 1.4cm 1.0cm, clip]{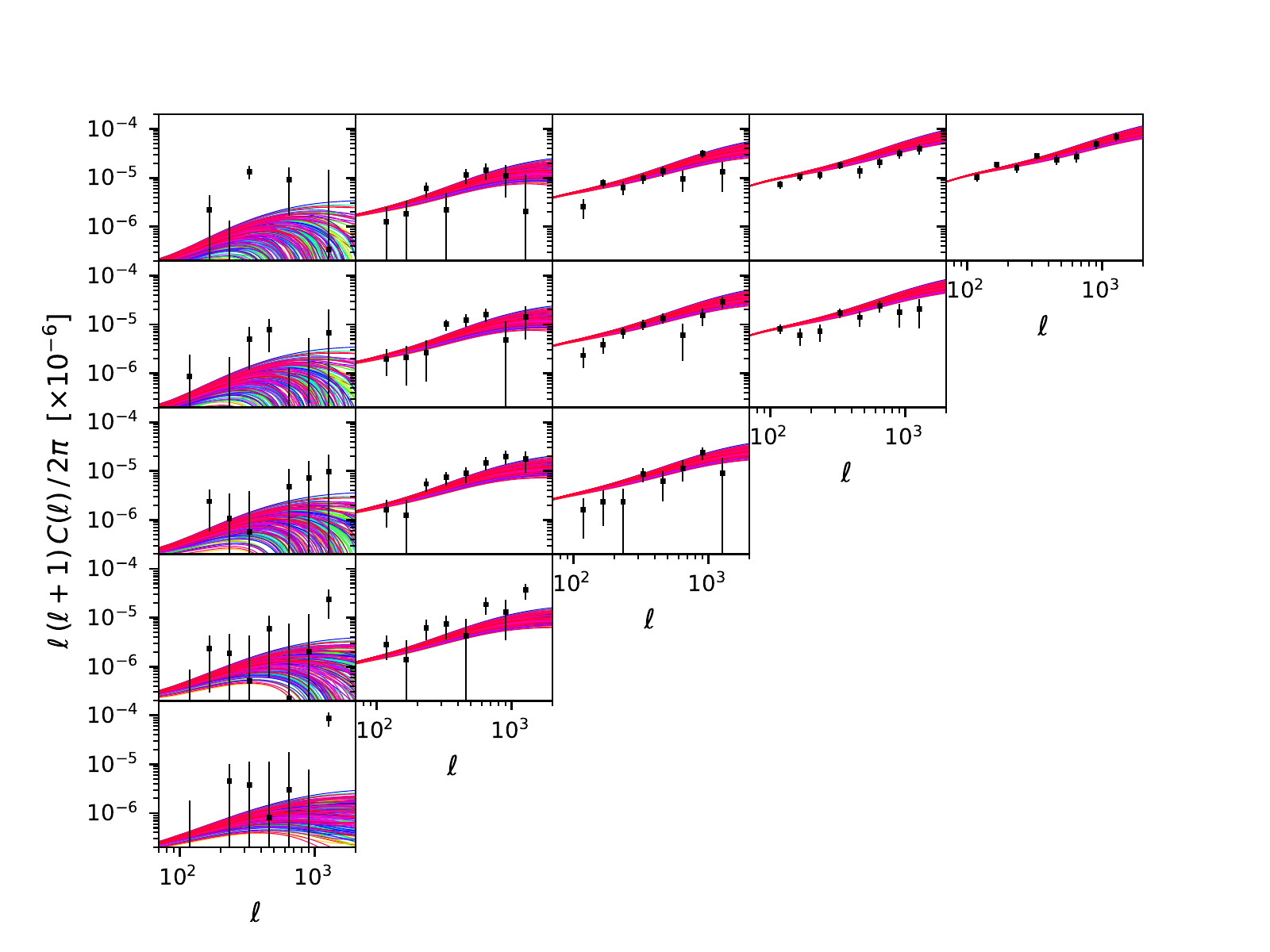}
\caption{Angular weak-lensing shear power spectra assuming a {\tt Planck} 2018 cosmology and randomly selected parameter values for the baryonification model (coloured lines). The baryonic parameters all lie within the prior range of the baryonic emulator \citep{Giri:2021qin} which provides a very conservative estimate for the current range of uncertainty due to baryonic feedback. The data points correspond to the band power from the {\tt KiDS}-1000 data release \citep{KiDS:2020suj}. Different panels show the auto- and cross-correlation spectra of the 5 different tomographic bins between $z=0$ (bottom) and 1.5 (top). }
\label{fig:cl_example}
\end{figure*}

\section{Data and analysis pipeline}\label{sec:data}
In this section we discuss all data products used for the analysis. This includes the weak-lensing shear spectra, the X-ray cluster fractions, and the stacked gas profiles from the kSZ observations. Next to the data, we also present the analysis pipeline to predict the theory signal corresponding to each observable.

\subsection{Weak-lensing shear power spectrum}
\label{sec:WL}
For the weak-lensing shear power spectrum, we use observations from the Kilo-Degree Survey ({\tt KiDS}) 1000 data release \citep{Kuijken:2019gsa}. The data contains the band power spectrum separated into 8 logarithmically spaced data points in the range $\ell\sim100-2000$ and split up in five tomographic bins between $z=0-1.5$ \citep{KiDS:2020suj}. The full data vector is shown as black symbols in Fig.~\ref{fig:cl_example}. The corresponding covariance matrix is published in \citet{Joachimi:2020abi}. The square-root of its diagonal values are shown as error-bars on the data points of Fig.~\ref{fig:cl_example}.

The relation between the band power ($C_{ij}^l$) and the angular power spectrum ($C_{ij}$) is given by
\begin{equation}
C_{ij}^l = \frac{1}{2\mathcal{N}_l}\int_0^{\infty} d\ell \ell W_l(\ell)C_{ij}(\ell),
\end{equation}
where $W_l(\ell)$ are window functions for the different data-bins $l=1-8$. Here we have assumed the B-mode angular power spectrum to be exactly zero. The normalisation factor is defined as
\begin{equation}
\mathcal{N}_l= \ln(\ell_{\rm up, l})-\ln(\ell_{\rm dn, l}),
\end{equation}
with the bin-ranges designated by $\ell_{\rm up, l}$ and $\ell_{\rm dn, l}$, respectively.

The band window functions are illustrated in Fig.~2 of \citet{Joachimi:2020abi}. The same reference also provides the exact functional form of each window. Qualitatively, all band windows above $\ell\sim 500$ resemble tophat filters, while for smaller multipole-values, they are closer to Gaussian filters. In practice, using band powers instead of the angular power spectrum for the theory pipeline only has a very minor effect on the resulting parameter contours.

\begin{figure*} 
\centering
\includegraphics[width=0.49\textwidth,trim=0.2cm 0.1cm 1.5cm 1.0cm, clip]{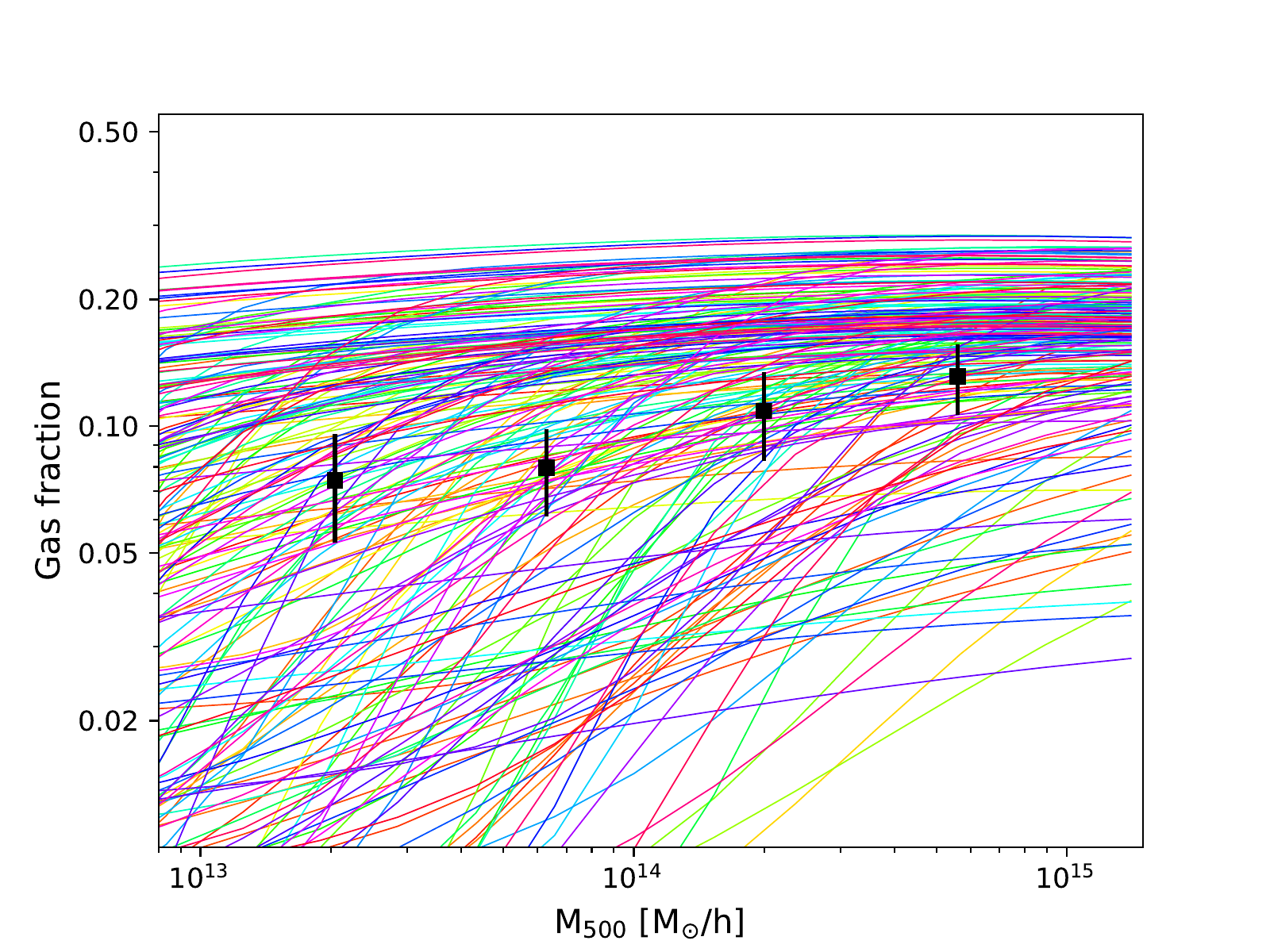}
\includegraphics[width=0.49\textwidth,trim=0.2cm 0.1cm 1.5cm 1.0cm, clip]{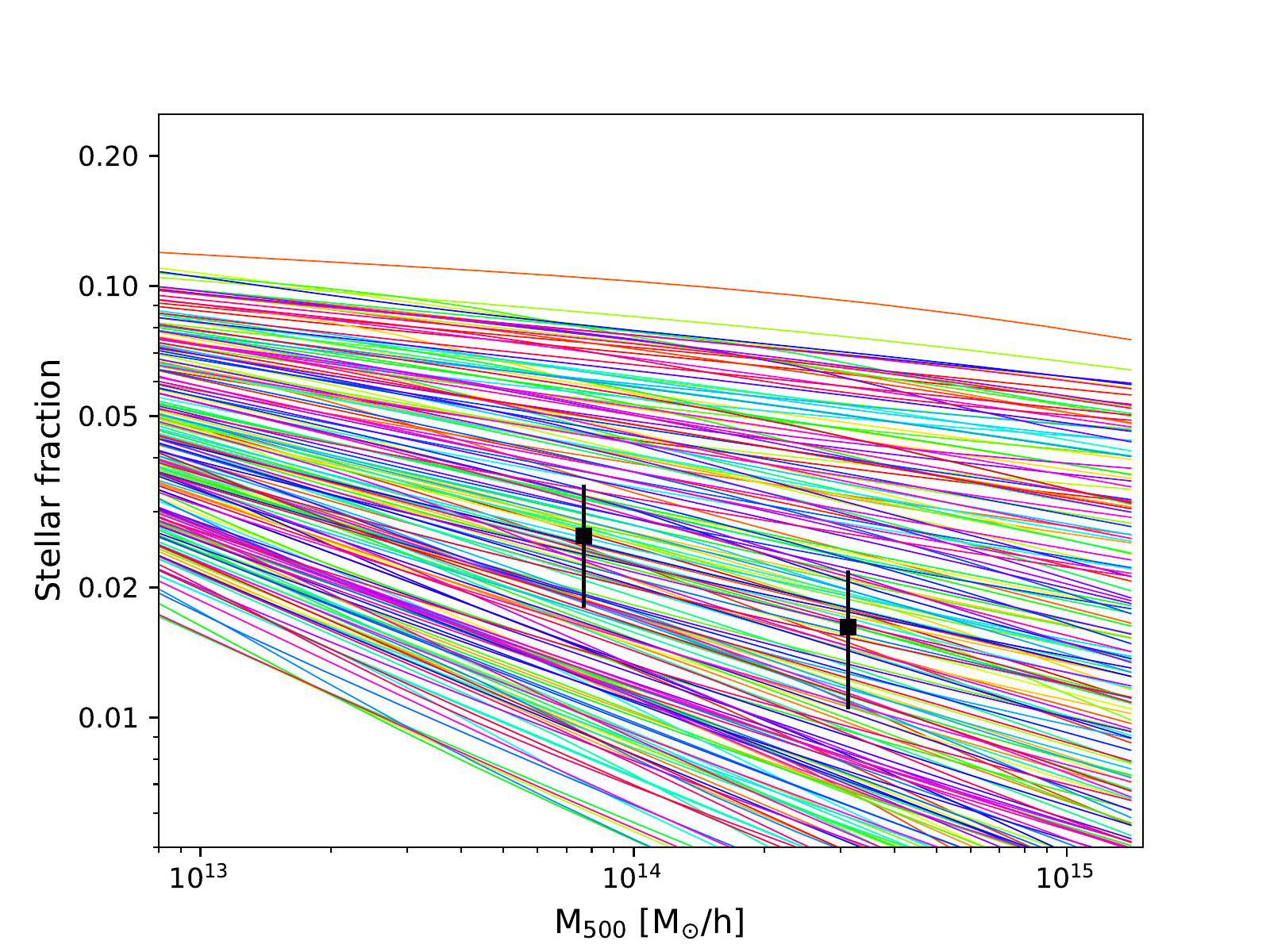}
\caption{Gas and stellar fractions assuming a {\tt Planck} cosmology and randomly selected parameter values for the baryonification model (coloured lines). The selected parameters are identical to the ones shown in Fig.~\ref{fig:cl_example}, randomly sampling the prior range of the baryonic emulator. The data points correspond to binned X-ray and optical measurements from individual galaxy groups and clusters \citep[see][]{Giri:2021qin}. The mass unit $M_{500}$ consists of the estimated X-ray mass at the radius $r_{500}$ assuming hydrostatic equilibrium.}
\label{fig:xray_example}
\end{figure*}

The angular power spectrum for the auto and cross-correlation is obtained using the Limber approximation
\begin{equation}
C_{ij}(\ell)=\int_0^{\chi_H}\frac{g_i(\chi)g_j(\chi)}{\chi^2}P_{\rm full-physics}\left(\frac{\ell}{\chi},z(\chi)\right)d\chi,
\end{equation}
with the co-moving distance $\chi$ going from 0 to the horizon $\chi_H$. The power spectrum $P_{\rm full-physics}(k,z)$ is given by
\begin{equation}
P_{\rm full-physics} = \mathcal{S}P_{\rm gravity-only},
\end{equation}
where $P_{\rm gravity-only}$ is obtained using the revised halofit predictor from \cite{Takahashi:2012em} and the multiplicative baryonic suppression factor $\mathcal{S}(k,z)$ is calculated with the baryonic emulator from \cite{Giri:2021qin}. Note that $\mathcal{S}$ depends on cosmology via the cosmic baryon fraction $f_b\equiv\Omega_b/\Omega_M$, which is emulated together with the seven parameters of the baryonification model\footnote{All other cosmological parameters have a negligible effect on the baryonic supression signal $\mathcal{S}$ \citep[as shown in][]{Schneider:2019snl}.}.

The lensing weights are given by
\begin{equation}
g_i(\chi) = \frac{3\Omega_m}{2}\left(\frac{H_0}{c}\right)^2\chi(1+z)\int_{\chi(z)}^{\chi_H}n_i(z)\frac{\chi(z')-\chi(z)}{\chi(z')}dz',
\end{equation}
where $n_i(z)$ is the galaxy distribution at redshift bin $i$. The galaxy redshift distributions are obtained form the {\tt KiDS} 1000 data release \cite[see Fig. 2 in][]{KiDS:2020suj}. The calculations of the weak-lensing band powers are performed using the {\tt PyCosmo} framework presented in \cite{Refregier:2017seh}.

Compared to the original analysis from {\tt KiDS} 1000, we simplify the analysis pipeline to some degree. First of all, we assume neutrinos to be mass-less. While it is known that at least two of the three neutrino flavour states have small but nonzero mass, it has been shown by \citet{Hildebrandt:2018yau} that massive neutrinos only lead to very minor changes of the posterior contours for the {\tt KiDS} 450 data. Second, we use the revised {\tt halofit} routine from \citet{Takahashi:2012em} together with the \citet{Eisenstein:1997ik} linear transfer function instead of the {\tt HMcode} \citep{Mead:2015yca} combined with {\tt Class} \citep{Blas:2011rf}. While \citet{Joachimi:2020abi} have reported a 0.3-$\sigma$ shift of the $S_8$ parameter between {\tt HMcode} and {\tt halofit}, we do not find such a large deviation between the two cases. Finally, we assume a fixed shift of the galaxy redshift distribution functions, following the findings of \cite[see their Table 1]{KiDS:2020suj}, but we do not add new nuisance parameters to vary the mean of the redshift distributions (as it is the case in the standard {\tt KiDS} analysis). Our approach accounts for the shift of the posterior contours due to errors in the mean tomographic redshift estimates, but it does not include the potential broadening of the posterior contours that may (or may not) be induced by additional nuisance parameters. Indeed, \citet{KiDS:2020suj} showed that shifting the mean of the redshift distributions back to their default positions leads to a small shift of $S_8$ (of order $0.3$-$\sigma$) while the size of the error remains unchanged. We therefore conclude that our approach (of accounting for the shifts of the mean redshift distributions without adding 5 more nuisance parameters) has no significant effect on the $S_8$ posterior contours.

In Fig.~\ref{fig:cl_example} we plot 200 angular power spectra with varying baryonic model parameters (and fixed cosmology to the {\tt Planck} values) as coloured lines. The baryonic parameters have thereby been assigned randomly within the prior ranges given by the baryonic emulator of \citet[see also Table~\ref{table:par_ranges}]{Giri:2021qin}. These prior ranges are selected to be significantly broader than all known results from hydrodynamical simulations. The spread of the coloured lines therefore provides a conservative estimate for the maximum uncertainty from baryonic feedback on the weak-lensing signal. 
Note that in Fig.~\ref{fig:cl_example}, the observed band power values (black symbols) are plotted together with the predicted angular power spectra (coloured lines). Although not exactly the same, both measures remain within a few percent from each other, a shift that is not visible in the plot.

\subsection{X-ray gas and stellar fractions}
As mentioned above, the baryonification model can be used to recover the baryonic suppression of the matter power spectrum based solely on information about the halo gas profiles \citep[see e.g.][]{Schneider:2018pfw,Arico:2020yyf,Giri:2021qin}. For example, it is possible with X-ray data to obtain detailed information about the gas content within the virial radius of galaxy groups and clusters. Furthermore, X-ray observations can be used to estimate the total halo mass (with the assumption of hydrostatic equilibrium). Combining weak-lensing data with X-ray observations can therefore help to provide information about baryonic effects and lead to improved constraints for cosmological parameters \citep[][]{Schneider:2019xpf}.

\begin{figure*} 
\centering
\includegraphics[width=0.49\textwidth,trim=0.2cm 0.1cm 1.5cm 1.0cm, clip]{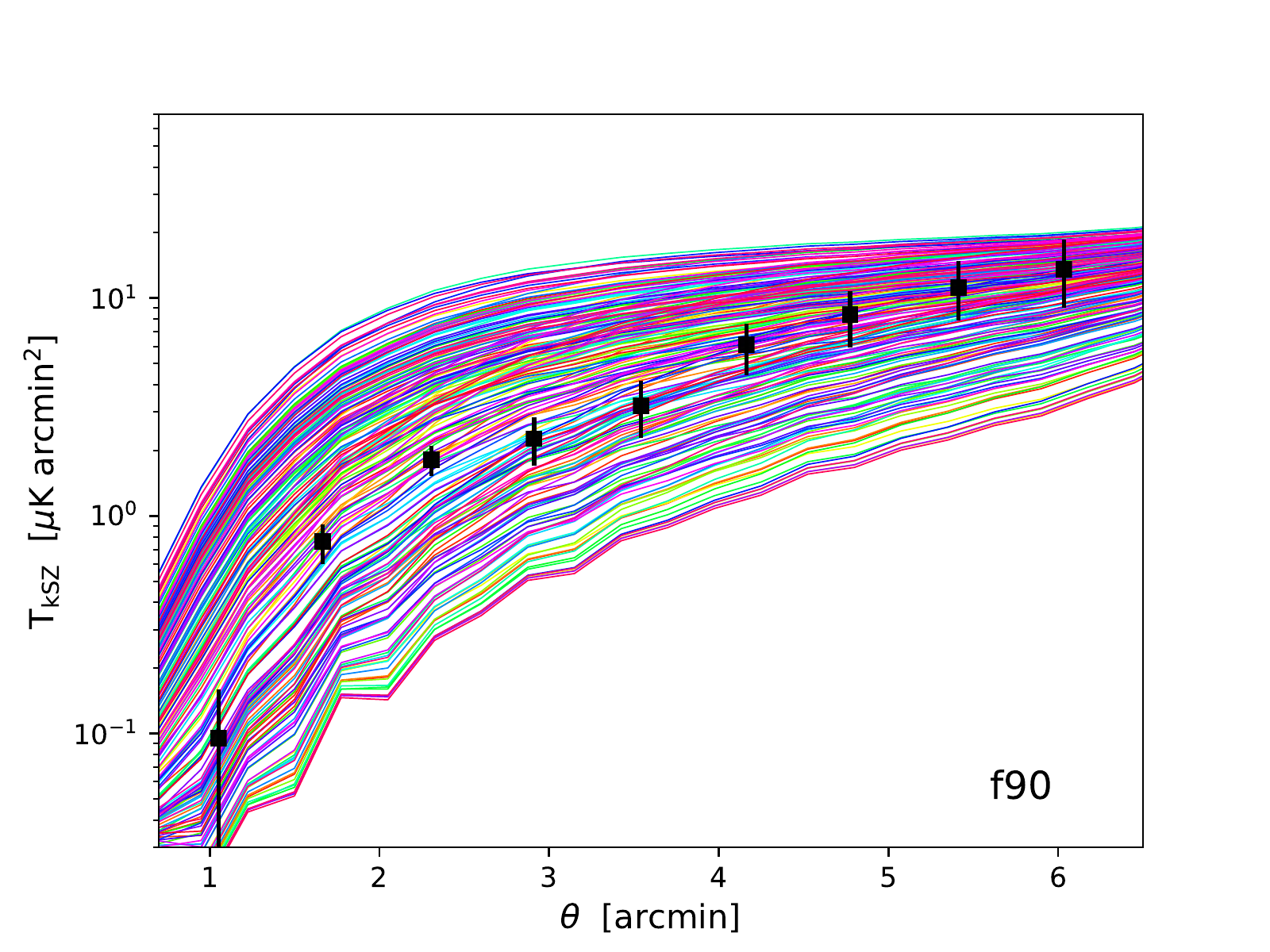}
\includegraphics[width=0.49\textwidth,trim=0.2cm 0.1cm 1.5cm 1.0cm, clip]{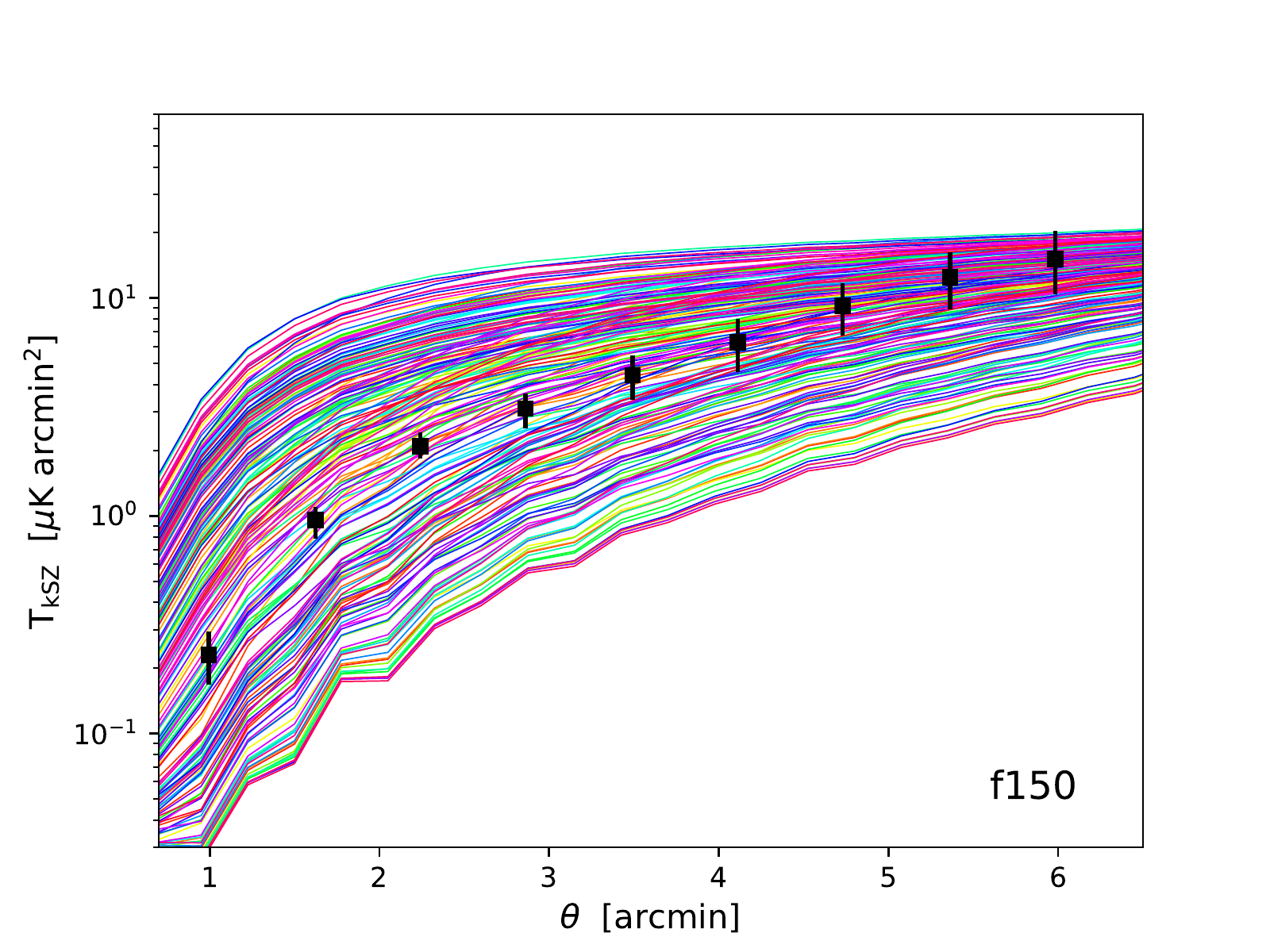}
\caption{Stacked kSZ profiles from {\tt ACT} at 98 GHz (f90, left-hand panel) and 150 GHz (f150, right-hand panel) at $z=0.55$ (black data points). The coloured lines correspond to the predicted kSZ profiles for a halo with $M_{200}=3\times10^{13}$ M$_{\odot}$ assuming a fixed {\tt Planck} cosmology and randomly varying parameters of the baryonification model. The predictions correspond to the same models shown in Fig.~\ref{fig:cl_example} and \ref{fig:xray_example}, providing a conservative estimate for the maximum variation due to baryonic feedback processes.}
\label{fig:SZ_example}
\end{figure*}

In this paper we will use the binned X-ray gas and stellar fractions presented in \cite{Giri:2021qin}. The data is built upon individual measurements of the gas and stellar content of galaxy groups and clusters from various different Refs.~\citep{Vikhlinin:2008cd,Gonzalez:2013awy,Sanderson:2013aaa,Lovisari:2015aaa,Kravtsov:2018aaa}. The binned data is shown as black symbols in Fig.~\ref{fig:xray_example} with the gas fraction to the left and the stellar fraction to the right. Both fractions show a clear trend as a function of halo mass ($M_{500}$). Galaxy groups and small clusters are lacking gas compared to large clusters, where the gas fraction approaches the mean cosmological fraction $f_g=\Omega_b/\Omega_m -f_{\rm star}$. For the stellar component, the trend is reversed with a rising total stellar fraction towards smaller halo masses.

For the prediction of the X-ray gas and stellar fractions, we use the stellar, gas and halo masses of the baryonification model, i.e.,
\begin{equation}\label{fraction}
f_{\chi} = \frac{M_{\chi}(r_{500})}{(1+b_{\rm hse})M_{\rm tot}(r_{500})},
\end{equation}
with $\chi=\lbrace{\rm gas},\, {\rm star}\rbrace$ and where $r_{500}$ is the radius corresponding to a mean halo density of 500 times the critical density of the universe. In Eq.~(\ref{fraction}) we have introduced the hydrostatic bias $b_{\rm hse}$ that accounts for deviations from hydrostatic equilibrium due to in-falling gas in the outskirts of haloes. We use $b_{\rm hse}$ as an additional model parameter assuming a Gaussian prior with a mean of 0.26 and a standard deviation of 0.07 \citep{Hurier:2017jkv}. This assumption is based on an analysis of CMB data and agrees well with other direct estimates from X-ray and lensing observations \citep[see e.g. Fig. 10 in][]{Salvati:2017rsn}.

The predicted gas and stellar fractions for different baryonic parameters (at fixed cosmology) are shown in Fig.~\ref{fig:xray_example} as coloured lines. The model parameters are the same than in Fig.~\ref{fig:cl_example} and have been selected randomly within the prior ranges provided in Table~\ref{table:par_ranges}. The potential of X-ray data to constrain baryonic parameters becomes evident when comparing the errors of the data points in Fig.~\ref{fig:xray_example} to the widely spread predictions from the baryonification model.

\subsection{Profiles from the kinematic Sunyaev-Zeldovich effect}
Another promising observable of the gas around haloes arises from the Sunyaev-Zeldovich effect \citep[SZ,][]{SZ1970, SZ1972}. The predominant process is the thermal SZ (tSZ), a spectral distortion of the CMB black-body spectrum due to inverse Compton scattering of the CMB photons by the hot free electrons in galaxy clusters, which is proportional to the integrated electron pressure along the line of sight. Very recently, \cite{Gatti:2021tjt} and \cite{Troster:2021gsz} have published first result of tSZ measurements cross-correlated with weak-lensing data from {\tt DES} and {\tt KiDS}, respectively. They have shown that such cross-correlation studies can be used to simultaneously constrain cosmology and baryonic feedback parameters.

In this paper we focus instead on the kSZ %kinematic Sunyaev-Zeldovich (kSZ) 
effect, consisting of an up-scattering of the CMB photons due to the bulk motion of the free electrons in galaxy groups and clusters. This is proportional to the integrated electron momentum along the line of sight, therefore, if the velocity is known, it provides a direct measurement of the electron density, as opposed to the tSZ which depends also on the electron temperature. %, and to the X-ray observable (surface brightness) which is proportional to the square of the density and depends on the temperature as well. (what did you mean by that?)
For this reason, the kSZ effect is a better probe of lower density environments compared to the tSZ and X-ray measurements, such as the regions further away from the halo centre and smaller halo masses. We use kSZ measurements from the Atacama Cosmology Telescope ({\tt ACT}) DR5  and {\tt Planck} \citep{AtacamaCosmologyTelescope:2020wtv, Amodeo:2020mmu} obtained for the CMASS galaxy sample from the Baryon Oscillation Spectroscopic Survey \citep[{\tt BOSS},][]{SDSS:2013qvl}, having average halo mass and redshift of $M_{200}=3\times10^{13}$ M$_{\odot}$ and  $z=0.55$.

The kSZ temperature measured within a disc of radius $\theta$ around a galaxy group or cluster is given by the relation
\begin{equation}\label{TkSZ}
T_{\rm kSZ}(\theta)=T_{\rm CMB} \left(\frac{v_{e,r}}{c}\right)\tau_{\rm gal}(\theta),
\end{equation}
where $\tau_{\rm gal}$ is the Thomson optical depth and $v_{e,r}$ is the free electron bulk line-of-sight velocity, approximated as $v_{e,r}=1.06\times10^{-3}c$ \citep{AtacamaCosmologyTelescope:2020wtv, Amodeo:2020mmu}. The optical depth is obtained via a line-of-sight integral over the electron number density profile, i.e.,
\begin{equation}
\tau_{\rm gal}(\theta) = 2\sigma_T\int_0^{\infty}n_{e}\left(\sqrt{l^2+d_A(z)^2\theta^2}\right)dl,
\end{equation}
with $\sigma_T$ being the Thomson scattering cross-section and $d_A(z)$ the angular diameter distance. The electron density $n_e(r)$ is related to the gas profile via
\begin{equation}
n_e(r) = \frac{(X_H+1)}{2m_{\rm amu}}\rho_{\rm gas}(r),
\end{equation}
with the hydrogen mass fraction $X_H=0.76$ and the atomic mass unit $m_{\rm amu}$. Note that the gas profile $\rho_{\rm gas}(r)$ is directly obtained from the baryonification model and contains both the one-halo and the two-halo term.

In order to compare the kSZ temperature to observations from the {\tt ACT} survey, we convolve Eq.~(\ref{TkSZ}) with the beam profile measured at 98 and 150 GHz (called f90 and f150). These beam profiles are assumed to be of Gaussian shape (with ${\rm FWHM}=2.1$ and $1.3$ arcmin for f90 and f150, respectively) which has shown to be a good approximation \cite[see Fig. 5 of][]{AtacamaCosmologyTelescope:2020wtv}. Finally, we smooth the data using a step-like filter function that is 1 within $\delta_d$, -1 between $\delta_d$ and $\sqrt{2}\delta_d$, and 0 otherwise \cite[see Eq. 9 of][]{AtacamaCosmologyTelescope:2020wtv}. Applying this filter is necessary to reproduce the aperture photometry filtering used for the data.

\begin{figure*} 
\centering
\includegraphics[width=0.99\textwidth,trim=0.2cm 0.1cm 0.9cm 0.2cm, clip]{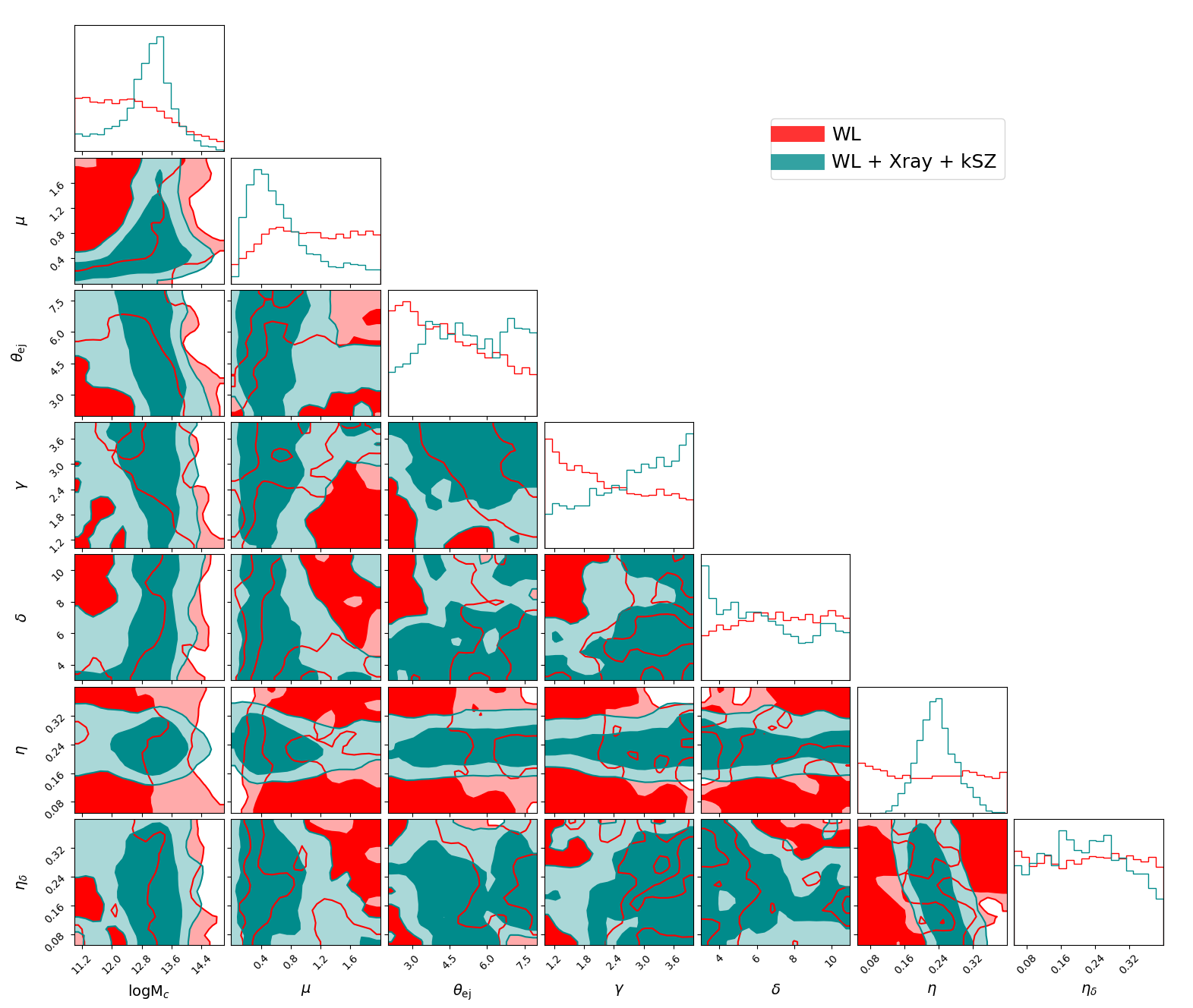}
\caption{Posterior contours of the baryonification model parameters at 68 and 95 percent confidence level. Results based on the weak-lensing (WL) band-power data from KiDS 1000 \citep{KiDS:2020suj} are shown in red colour. Results from the combined data set including KiDS 1000 WL data, X-ray gas (and stellar) fractions from \citet{Giri:2021qin}, and stacked kSZ profiles from ACT \citep{AtacamaCosmologyTelescope:2020wtv} are shown in dark cyan colour. Posterior contours from the WL+Xray and WL+kSZ data combinations are provided in Fig.~\ref{fig:bfc_constraints2}.}
\label{fig:bfc_constraints}
\end{figure*}

In Fig.~\ref{fig:SZ_example}, we show the predicted kSZ profiles assuming a host halo of $M_{200}=3\times10^{13}$ M$_{\odot}$ at $z=0.55$ (coloured lines). The different models are characterised by randomly varying parameters of the baryonification model (at fixed cosmology). They consist of the same models shown in Fig.~\ref{fig:cl_example} and \ref{fig:xray_example}, covering the full prior range of the baryonic parameters given in Table~\ref{table:par_ranges} (which is a conservative estimate for the maximum variation of the baryonic feedback processes). Note that although the kSZ signal is independent of frequency, distinct profiles are measured due to the different beams at 98 GHz (f90) and 150 GHz (f150), shown in the left- and right-hand panels, respectively.

The black data points in Fig.~\ref{fig:SZ_example} show the stacked kSZ signal from the Atacama Cosmology Telescope ({\tt ACT}) at 98 and 150 GHz \citep{AtacamaCosmologyTelescope:2020wtv}. The stacking was performed by cross-correlating the {\tt ACT} signal with the CMASS galaxy catalogue. The corresponding galaxies that appear in both data-sets have a mean stellar mass of $M_{\rm star}\sim 3\times10^{11}$ M$_{\odot}$ which corresponds to a halo mass of $M_{\rm 200}\sim 3\times10^{13}$ M$_{\odot}$ assuming the abundance matching estimate of \cite{Kravtsov:2018aaa}.

For the inference analysis performed in this paper, note that we do not assume a fixed value for the halo mass but we keep $M_{200}$ as a free model parameter (within the prior limits given in Table~\ref{table:par_ranges}). Furthermore, we account for the fact the the error-bars of the stacked profiles from ACT are strongly correlated by including the error covariances \citep[see Fig. 7 of][]{AtacamaCosmologyTelescope:2020wtv}.

Fig.~\ref{fig:SZ_example} shows that the variations in the shape and amplitude of kSZ profiles from different feedback parameters widely exceeds the size of the error-bars from the ACT observations. This illustrates the great potential of kSZ data to provide constraints on the strength of baryonic feedback.

\begin{figure*} 
\centering
\includegraphics[width=0.49\textwidth,trim=0.2cm 0.1cm 1.5cm 1.0cm, clip]{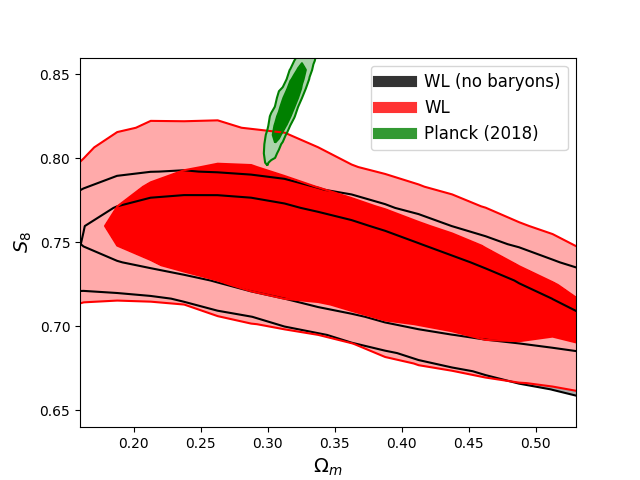}
\includegraphics[width=0.49\textwidth,trim=0.2cm 0.1cm 1.5cm 1.0cm, clip]{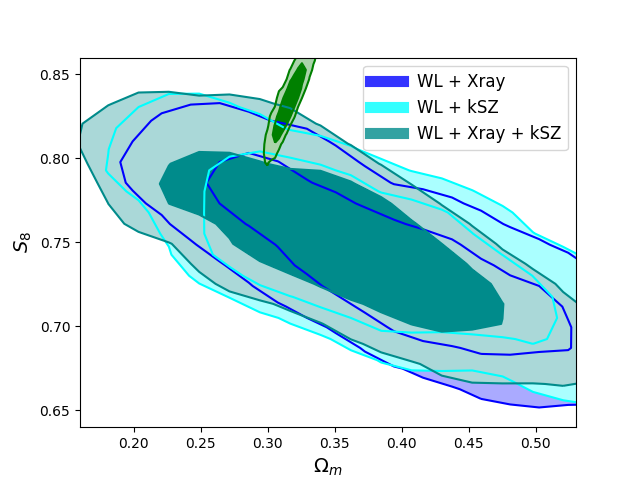}
\caption{Constraints on the cosmological parameters $\sigma_8$ and $\Omega_m$ (left-hand panel) and on the combined $S_8=\sigma_8\sqrt{\Omega_m/0.3}$ parameter (right-hand panel). The black and red posterior contours show results from weak-lensing data alone, assuming no baryonic effects and marginalising over all baryonification parameters. Results from the combined analysis including weak-lensing and X-ray gas fraction and kSZ profiles data are again shown in blue and cyan. The green contours correspond to CMB estimates from {\tt Planck}. All posteriors are shown at 68 and 95 percent confidence level, respectively.}
\label{fig:cos_constraints}
\end{figure*}

\section{Results}\label{sec:results}
Based on the data and the theory-pipeline discussed above, we now perform an inference analysis using the Markov Chain Monte Carlo (MCMC) sampling method. We thereby simultaneously vary five cosmological parameters ($\Omega_m$, $\Omega_b$, $\sigma_8$, $n_s$, $h_0$), seven baryonic parameters ($M_c$, $\mu$, $\theta_{\rm ej}$, $\delta$, $\gamma$, $\eta$, $\eta_{\delta}$), and one parameter accounting for intrinsic alignment ($A_{\rm IA}$). Whenever we include X-ray and/or kSZ data, we furthermore vary the hydrostatic bias parameter ($b_{\rm hse}$) and/or the halo mass parameter of the kSZ profile ($M_{200}$). A list of all parameters including their prior ranges are provided in Table~\ref{table:par_ranges}.

As a first step, we produce MCMC samples based only on the weak-lensing band power from the {\tt KiDS}-1000 data release \citep{KiDS:2020suj}. We investigate two scenarios, one where we ignore all baryonic effects and one where we include the full baryonification model. Note that the former is only added to illustrate the effect of  baryonic feedback on the posterior contours and should not be regarded as physically viable scenario.

As a further step, we combine the weak-lensing band-power data with the mean X-ray cluster gas (and stellar) fractions \citep{Giri:2021qin} and with the stacked kSZ profiles at 98 and 150 GHz \citep{AtacamaCosmologyTelescope:2020wtv}. For these scenarios, we only investigate models including the full baryonification model, as the data is primarily added to provide further constraints on the baryonic parameters.

In the following sections, we discuss the outcome of the MCMC analyses, focusing first on the baryonic and then on the cosmological parameters. Finally, we also provide a observation-based estimate for the baryonic suppression of the matter power spectrum, and we compare it to predictions from the most recent hydrodynamical simulations. 

\begin{table}
\centering
\begin{tabular}{l c c} 
% \hline \\
 Parameter name & Acronym & Prior range \\ 
 \hline \hline
 Matter abundance & $\Omega_m$ & [0.1, 0.6]\\
 Baryon abundance & $\Omega_b$ & [0.04, 0.06]\\
 Clustering amplitude & $\sigma_8$ & [0.3, 1.3]\\
 Hubble parameter & $H_0$ & [0.64, 0.82]\\
 Spectral index & $n_s$ & [0.84, 1.10]\\
 \hline
 Intrinsic alignment parameter & $A_{\rm IA}$ & [0, 2] \\
 \hline
 First gas parameter (Eq. \ref{rhogas}) & $\mathrm{log_{10}} M_\mathrm{c}$ & [11, 15]\\ 
 Second gas parameter (Eq. \ref{rhogas}) & $\mu$ & [0, 2] \\
 Third gas parameter (Eq. \ref{rhogas}) & $\theta_\mathrm{ej}$ & [2, 8]\\
 Fourth gas parameter (Eq. \ref{rhogas})& $\gamma$ & [1, 4] \\
 Fifth gas parameter (Eq. \ref{rhogas})& $\delta$ & [3, 11] \\
 First stellar parameter (Eq. \ref{fstar})& $\eta$ & [0.05, 0.4] \\
 Second stellar parameter (Eq. \ref{fstar})& $\eta_{\delta}$ & [0.05, 0.4] \\
 \hline
 Hydrostatic mass bias (Eq. \ref{fraction}) & $b_{\rm hse}$ & $0.26\pm0.07$\\
 \hline
 Total halo mass (for kSZ profiles) & $M_{200}$ & [1e13, 4e13]\\
 \hline\hline
\end{tabular}
\caption{List of MCMC parameters separated into cosmological, intrinsic alignment, baryonic, hydrostatic bias, and kSZ halo mass parameters (top to bottom). We use a Gaussian prior for $b_{\rm hse}$ and flat priors for all other parameters.}
\label{table:par_ranges}
\end{table}

\subsection{Constraining the baryonification model}
As a first result, we provide constraints for the baryonification model parameters. In Fig.~\ref{fig:bfc_constraints}, the posterior contours from the {\tt KiDS}-1000 WL band-power alone (red) are compared to the contours from the combined WL + Xray + kSZ data-set (dark cyan). The results from the data combinations WL + Xray and WL + kSZ are shown in Fig.~\ref{fig:bfc_constraints2} of Appendix~\ref{app:bfcconstraints}.

Fig.~\ref{fig:bfc_constraints} shows that the WL band-power data alone is unable to constrain the baryonic model parameters very well. Only the $M_c$, $\theta_{\rm ej}$, and $\gamma$ parameters show a slight preference for small values without, however, excluding larger values either. The other parameters yield fairly flat distributions without preference for any particular value within the prior ranges.

Not surprisingly, the combined WL + Xray + kSZ data vector leads to stronger constraints on the baryonic  parameters. Especially $M_c$, $\mu$, and $\eta$ are well constrained, while $\gamma$ and (to a lesser degree) $\theta_{\rm ej}$ are pushed towards the upper end of the prior range. The remaining parameters  $\delta$ and $\eta_{\delta}$ stay relatively unconstrained, with a slight preferences for values at the lower end and at the centre of the prior range, respectively. A comparison with Fig.~\ref{fig:bfc_constraints2} reveals that the constraints on the stellar parameters $\eta$ and $\eta_{\delta}$ are entirely driven by the X-ray fractions that contain information about the stellar abundance in haloes. The posteriors of the gas parameters $M_c$, $\mu$, and $\delta$, on the other hand, are a product of combining both X-ray and kSZ data.

Fig.~\ref{fig:bfc_constraints} illustrates the potential of current and future gas observations to constrain baryonic feedback effects relevant for cosmology. In the future, combining e.g observations from {\tt Euclid} \citep{Laureijs:2011gra} with X-ray data from {\tt eROSITA} \citep{Merloni:2012uf} and SZ measurements from {\tt CMB-S4} \citep{CMB-S4:2016ple} will make it possible to further decrease the current uncertainties related to the baryonification model. In \citet{Schneider:2019xpf} it is shown that combining {\tt Euclid} weak-lensing observations with X-ray data from {\tt eROSITA} will make baryonification model errors shrink by more than an order of magnitude. As a consequence, we expect baryonic nuisance parameters to not substantially degrade cosmological parameter estimates, provided external data from X-ray or the SZ effect are included.

\subsection{Constraining cosmological parameters}
Arguably the most intriguing result from the recent weak-lensing surveys {\tt CFHTLens}, {\tt KiDS}, {\tt DES}, and {\tt HSC} is the mild yet persistent tension with the matter abundance ($\Omega_m$) and clustering amplitude ($\sigma_8$) measured by the CMB satellite {\tt Planck}. The tension is usually quantified in terms of the combined $S_8$ parameter, defined as
\begin{equation}\label{S8}
S_8\equiv\sigma_8\sqrt{\Omega_m/0.3}.
\end{equation}
In this section, we investigate the $S_8$-tension in connection with the baryonic suppression effects.

The left-hand panel of Fig.~\ref{fig:cos_constraints} illustrates the $\Omega_m$ and $S_8$ posterior contours (at 68 and 95 percent level) based on the WL band power data from {\tt KiDS} 1000 \citep{KiDS:2020suj}. The black contours correspond to the case where baryonic effects are ignored in the prediction pipeline (which means that the baryonic power suppression function $\mathcal{S}\equiv 1$ at all scales and redshifts). The red contours correspond to the model where all 7 parameters of the baryonification model are varied (and marginalised over). In green we show the contours obtained from the TT, TE, EE + lowE data combination of the {\tt Planck} legacy analysis \citep{Planck:2018vyg}. The plot confirms the tension between the WL shear and the CMB measurements highlighted by the {\tt KiDS} science team in \citet{KiDS:2020suj}. It also illustrates the need for a realistic modelling of baryonic feedback. Including baryonic nuisance parameters leads to a broadening of the contours in the $S_8$ direction, thereby alleviating the $S_8$-tension with the CMB.

\begin{figure*} 
\centering
\includegraphics[width=0.49\textwidth,trim=0.3cm 0.1cm 0.5cm 1.0cm, clip]{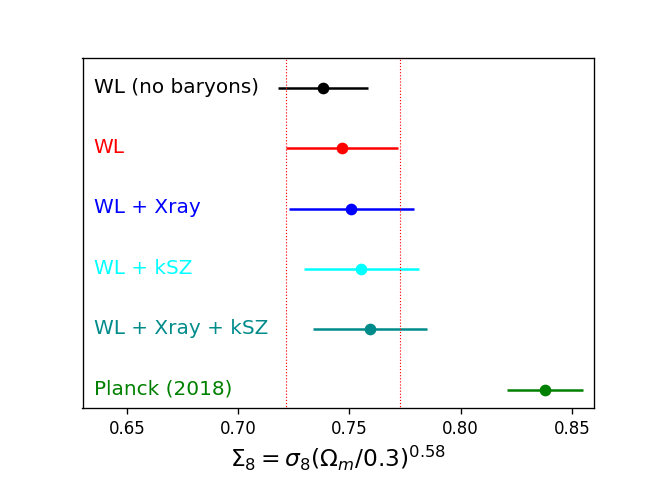}
\includegraphics[width=0.49\textwidth,trim=0.3cm 0.1cm 0.5cm 1.0cm, clip]{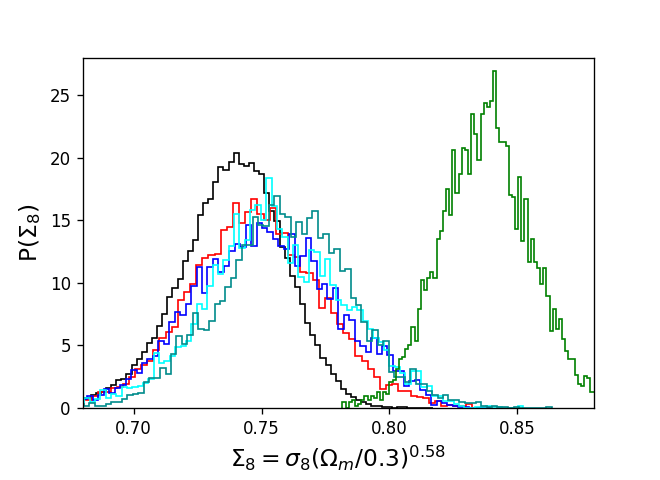}
\caption{Posterior distribution of the combined $\Sigma_8=\sigma_8(\Omega_m/0.3)^{0.58}$ parameter which is orthogonal to the $\Omega_m$-$\sigma_8$ degeneracy line. The left- and right-hand panels show the maximum likelihoods with 68 percent errors and the full distributions, respectively. The results based on the WL data alone are shown in grey and red (without and with baryonic suppression effect), from the WL + Xray and WL + kSZ data in blue and cyan, and from WL + Xray + kSZ in dark cyan.}
\label{fig:cos_constraints2}
\end{figure*}

In the right-hand panel of Fig.~\ref{fig:cos_constraints} we show the $\Omega_m$ and $S_8$ posterior contours for the cases where the WL band-power data is combined with observations from X-ray and the kSZ effect. In blue and cyan we show the results from the WL + Xray and the WL + kSZ data combination, respectively. They remain similar to the results from the WL only case (red), but are slightly shifted in the direction of the {\tt Planck} results. The same is true for the full WL + X-ray + kSZ data combination shown by the dark dark-cyan contours. We conclude that, while baryonic feedback visibly affects the posterior contours in the $\Omega_m$-$S_8$ plane, it is unable to fully solve the $S_8$ tension. This verdict based on WL shear is in qualitative agreement with the recent findings from galaxy-galaxy lensing \citep{Leauthaud:2016jdb,Lange:2020mnl,Amodeo:2020mmu}.

The original $S_8$ parameter defined in Eq.~(\ref{S8}) is not perfectly orthogonal to the degeneracy trend between the $\Omega_m$ and $\sigma_8$ parameters. A more optimal match is obtained by the parameter
\begin{equation}\label{Sigma8}
\Sigma_8\equiv\sigma_8\left(\Omega_m/0.3\right)^{\alpha},
\end{equation}
with best-fitting value $\alpha=0.58$ \citep{KiDS:2020suj}. For an approximate, single-parameter comparison of the tension between WL shear and the CMB data, it is therefore advisable to use $\Sigma_8$ instead of the original $S_8$ parameter.

In Fig.~\ref{fig:cos_constraints2} we illustrate the one dimensional constraints on the $\Sigma_8$ parameter. While the left-hand panel shows the maximum likelihood values with error-bars displaying the 68 percent confidence ranges, the right-hand panel provides the full posterior distributions. The colours are the same than in the previous plots (WL only results with and without baryonic effects in red and black, WL + Xray results in blue, WL + kSZ results in cyan, WL + Xray + kSZ results in dark cyan, and CMB results in green).

Assuming Gaussian distributions (which is not an unreasonable assumption, see right-hand panel of Fig.~\ref{fig:cos_constraints2}), we apply a simple measure to quantify the tension between the different data-sets given by $|\mu_A-\mu_B|/( \sigma_A+\sigma_B)^{1/2}$, where $\lbrace{\mu_A, \mu_B\rbrace}$ correspond to the means and $\lbrace \sigma_A,\sigma_B\rbrace$ to the variances of the distributions $\lbrace{ A, B\rbrace}$. With this measure, we find the WL data alone to be in 3.75 $\sigma$ tension with the {\tt Planck} measurement if baryonic effects are ignored. This tension is reduced to 2.94 $\sigma$ if baryonic effects are included and marginalised over. For the WL + Xray and WL + kSZ data combinations, we find a tension of 2.63 and 2.57 $\sigma$, respectively. Finally, the tension of the full WL + Xray + kSZ data set with the CMB observations from {\tt Planck} is also at 2.57 $\sigma$.

The comparison of the $\Sigma_8$ posteriors shows that ignoring baryonic feedback in the modelling pipeline leads to a bias of about half a standard deviation compared to the case where baryonic effects are included and marginalised over. Including additional information from X-ray and kSZ observations  pushes the distribution towards models with strong feedback effects, further reducing the tension with {\tt Planck} by about half a standard deviation. We again conclude that, while baryonic feedback models may induce a slight reduction of the tension between WL shear data and CMB observations, they cannot fully reconcile the two data-sets.

Note again that for the analysis presented above, we use a different pipeline than the {\tt KiDS} science team and we therefore obtain slightly different results than \citet{KiDS:2020suj}. A comparison at the level of the $\Omega_m$ and $S_8$ contours is provided in Appendix.~\ref{app:comparisonKiDS}.

\subsection{Constraining the matter power spectrum}
In this section we take a closer look at the baryonic suppression signal of the matter power spectrum. While current hydrodynamical simulations largely agree on the spoon-like shape of this suppression, there is a considerable scatter between the different predictions from simulations \citep[e.g.][]{Chisari:2019tus, Huang:2018wpy}. The differences are a result of the various feedback implementation and calibration strategies.

\cite{Giri:2021qin} has shown that the baryonification method is able to recover the power suppression from baryonic effects based solely on observed properties of the halo profiles. It is therefore possible to take advantage of the X-ray, kSZ, and (via the Limber approximation) the WL band power observations to predict the true amplitude and shape of the baryonic power suppression \citep[a method also suggested in][]{Arico:2019ykw}. This information can then be used to assess the validity of hydrodynamical simulations or to exclude biases affecting other cosmological probes based on the power spectrum (such as e.g the galaxy correlation).

\begin{figure*} 
\centering
\includegraphics[width=0.99\textwidth,trim=1.9cm 1.0cm 2.1cm 0.9cm, clip]{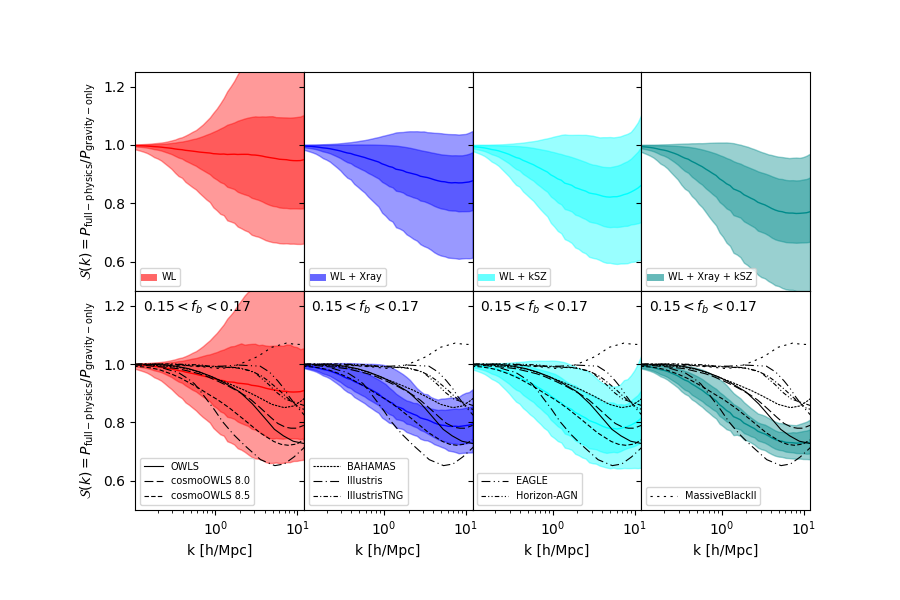}
\caption{Constraints on the baryonic suppression of the matter power spectrum from weak-lensing (WL) alone, WL plus X-ray fractions (Xray), WL plus stacked profiles from the kSZ, and the full data vector WL + Xray + kSZ (from left to right). All contours are shown at 68 and 95 percent confidence level. \emph{Top-panels:} All cosmological, baryonic, and intrinsic alignment parameters are marginalised over. \emph{Bottom-panels:} Only models with a cosmic baryon fraction ($f_b\equiv \Omega_b/\Omega_m$) inside the range $0.15<f_b<0.17$ are included into the analysis. This range comfortably includes all {\tt WMAP} and {\tt Planck} cosmologies and allows for a direct comparison with recent hydrodynamical simulations which are all based on either {\tt Planck} or {\tt WMAP} (black lines).}
\label{fig:ps_constraints}
\end{figure*}

In the top-panels of Fig.~\ref{fig:ps_constraints} we plot the constraints on the baryonic power suppression signal at 68 and 95 percent confidence level. The leftmost panel shows the results from the WL band power alone (red), the second and third panels illustrate the results from the combined WL + Xray (blue) and WL + kSZ (cyan) data sets, and the rightmost panel corresponds to the combined WL + Xray + kSZ data vector (dark cyan). Note that these constraints include uncertainties related to baryonic, intrinsic-alignment, and cosmological parameters.

The constraints from the WL band power alone (first top-panel of Fig.~\ref{fig:ps_constraints}) are rather loose. While they hint towards a suppression of the power spectrum, they do not exclude cases where baryons induce a net boost of power instead of a downturn. Considering wave-modes below $k=10$ h/Mpc, the smallest and largest values of the suppression factor at 95 percent confidence level are $\mathcal{S}=0.66$ and $\mathcal{S}=1.39$, respectively. Note that these findings are in qualitative agreement with \cite{Yoon:2020bop}

The combined WL and X-ray data set (second top-panel of Fig.~\ref{fig:ps_constraints}) leads to a stronger baryonic suppression signal compared to the WL observations alone. Models with a net power boost (and no sign of suppression) can be excluded at 68 (but not at 95) percent confidence. The smallest wave-mode where the baryonic feedback yields a suppression beyond one percent lies between $k_{\rm min}=0.12$ and $2.10$ h/Mpc. The largest amplitude of the suppression signal (where $\mathcal{S}$ is at its minimum) lies in the range $\mathcal{S}_{\rm min}=0.77-0.96$. Both measurements are provided at the 68 percent confidence level.

The results based on the combined WL and kSZ data (third top-panel of Fig.~\ref{fig:ps_constraints}) are very similar to the WL + X-ray case. Models with a net power boost are excluded at the 68 (but not at 95) percent level. The smallest wave-mode where the baryonic suppression signal exceeds one percent lies within the range $k_{\rm min}=0.10-0.61$ h/Mpc. The largest amplitude of the suppression signal is constrained to be within $\mathcal{S}_{\rm min}=0.72-0.92$. As before, these measures correspond to the 68 percent confidence level.

Finally, the right-most top-panel of Fig.~\ref{fig:ps_constraints} shows the baryonic power suppression from the combined WL + Xray + kSZ data. The resulting signal excludes a net power boost (no suppression) at more than 95 percent confidence. The amplitude of the suppression $\mathcal{S}$ exceeds one percent above $k_{\rm min}=0.1-0.45$ h/Mpc and reaches a minimum value of $\mathcal{S}_{\rm min}=0.67-0.88$ at $k\sim 7$ h/Mpc. Both these numbers are again given at the 68 percent confidence level.

For now, we have provided estimates of the baryonic power suppression assuming no prior knowledge on cosmology. Fixing the cosmological parameters will inevitably reduce the uncertainties illustrated in Fig.~\ref{fig:ps_constraints}. This is especially true regarding the cosmic baryon fraction $f_b=\Omega_b/\Omega_m$, which has been shown to directly affect the baryonic suppression signal. Other cosmological parameters have a negligible effect on the amplitude of the suppression $\mathcal{S}$ \citep[see][]{Schneider:2019snl}. 

Now we multiply the posterior distribution with a uniform prior of $0.15<f_b<0.17$. The prior range is selected so that it comfortably includes the best fitting cosmologies of {\tt WMAP} and {\tt Planck}. This allows us to obtain results that can be directly compared to various hydrodynamical simulations which are all based on either {\tt WMAP} or {\tt Planck} cosmologies. Note that, in principle, the WL + Xray + kSZ data-set could be combined with the full data vector from {\tt Planck} (instead of only assuming a CMB-informed prior on $f_b$). We refrain from doing so because the internal tension between these two data sets would push the parameter values towards very strong feedback models potentially biasing our conclusions.

The resulting estimates for the baryonic power suppression are plotted in the bottom-panels of Fig.~\ref{fig:ps_constraints}. As expected, adding a prior on the cosmic baryon fraction leads to a tightening of the uncertainty contours. The effect is most prominent for the WL + Xray and the full WL + Xray + kSZ data combination. The reason for this tightening is the data from the X-ray gas fractions which directly depends on the $f_b$ parameter. Once $f_b$ is constricted to a narrow range, the gas fractions provide much stronger constraints on the baryonification model.

Based on the data shown in the bottom-panels of Fig.~\ref{fig:ps_constraints}, we now quantify the constraints on the baryonic suppression signal in terms of scale and amplitude. For the smallest wave-mode ($k_{\rm min}$) where $\mathcal{S}$ falls below 0.99, we obtain $k_{\rm min}=0.11-0.32$  ($0.1-1.2$) h/Mpc for WL + Xray, $k_{\rm min}=0.11-0.53$ ($0.1-2.0$) h/Mpc for WL + kSZ, and $k_{\rm min}=0.11-0.22$ ($0.1-0.45$) h/Mpc for WL + Xray + kSZ at the 68 (95) percent confidence level. For the minimum value of the baryonic suppression signal ($\mathcal{S}_{\rm min}$) we get $\mathcal{S}_{\rm min}=0.73-0.84$ ($0.69-0.88$) for  WL + Xray, $\mathcal{S}_{\rm min}=0.70-0.86$ ($0.64-0.92$) for  WL + kSZ, and $\mathcal{S}_{\rm min}=0.69-0.76$ ($0.67-0.79$) for the combined WL + Xray + kSZ data set, again at 68 (95) percent confidence level. Finally, we notice that the $f_b$-prior on the cosmic baryon fraction has only little influence on the results from the WL only case (dark red, left-hand panel). While a slight tightening of the contours is visible, it is not enough to draw conclusions that are significantly different from the ones obtained without further assumptions regarding the cosmic baryon fraction.

The black lines in the bottom-panels of Fig.~\ref{fig:ps_constraints} correspond to the predicted baryonic power suppression signal from various hydrodynamical simulations. We show the results from {\tt OWLS} \citep[solid,][]{Schaye:2010aaa, vanDaalen:2011xb}, {\tt cosmo-OWLS} 8.0 and 8.5 \citep[long dashed and short dashed,][]{LeBrun:2014aaa,Mummery:2017lcn}, {\tt BAHAMAS} \citep[dotted,][]{McCarthy:2016mry}, {\tt Illustris} \citep[long dash-dotted,][]{Vogelsberger:2014dza}, {\tt IllustrisTNG} \citep[short dash-dotted,][]{Springel:2017tpz},  {\tt EAGLE} \citep[long dash-double-dotted,][]{Schaye:2014tpa,Hellwing:2016ucy}, {\tt H-AGN} \citep[short dash-double-dotted,][]{Dubois:2014lxa,Chisari:2018prw}, and {\tt MassiveBlackII} \citep[long dotted,][]{Khandai:2014gta}. All simulations except {\tt Illustris} predict a baryonic power suppression signal in agreement with the WL only data from {\tt KiDS}. The situation is very different when compared to the WL + Xray data set, where the power spectra of all simulations except {\tt OWLS}, {\tt cosmo-OWLS}, and {\tt BAHAMAS} lie outside of the 95 percent confidence region, at least for some of the wave-modes between 0.1 and 10 h/Mpc. Form the simulations that disagree, all except {\tt Illustris} show milder suppression signals than suggested by the WL + Xray data. The best match is obtained by {\tt cosmo-OWLS} 8.5 which stays within the 68 percent confidence region over all scales investigated.

More simulations agree with the analysis based on the less constraining WL + kSZ data set. While the {\tt OWLS}, {\tt cosmo-OWLS}, and {\tt BAHAMAS} power spectra stay within the 68 percent confidence region over all scales, the majority of simulations stay within the 95 percent region. Exceptions are {\tt EAGLE}, {\tt H-AGN}, and {\tt MassiveBlackII} that yield a significantly milder suppression signal than suggested by the WL + kSZ data.

The full WL + Xray + kSZ data set provides the most stringent constraints of the baryonic power suppression signal. Only the power spectrum of the {\tt cosmo-OWLS} 8.5 simulation remains in full agreement with the data. All other simulations predict power spectra that lie outside the 95 percent confidence range at least for certain $k$-values.

It is interesting to note that the {\tt BAHAMAS} simulation \citep{McCarthy:2016mry}, which has been specifically calibrated to data from X-ray gas fractions, lies outside of the 68 percent range of the WL + X-ray case. We speculate that this could be due to the hydrostatic mass bias from \citep{Hurier:2017jkv} that has been obtained from CMB observations. This direct estimate is somewhat larger than what is typically found in hydrodynamical simulations and could lead to an increase in amplitude of the baryonic suppression signal. Another possibility are resolution effects leading to an artificial increase of the {\tt BAHAMAS} results. Such an effect has been reported in \citet{vanDaalen:2019pst} and could, at least somewhat, alleviate the discrepancy sown in the bottom-panels of Fig.~\ref{fig:ps_constraints}. 

We conclude that the WL band-power data from {\tt KiDS} 1000 alone provides only weak constraining power on the baryonic feedback effects. This is in qualitative agreement with previous work \citep[e.g.][]{Fluri:2019qtp,DES:2020rmk,Yoon:2020bop, Troster:2021gsz}. For example, \citet{DES:2020rmk} find that the {\tt DES} year 1 combined with {\tt Planck} data is able to exclude both the dark-matter only case and the most extreme hydrodynamical simulations ({\tt Illustris} and {\tt cosmo-OWLS} 8.5) at the 68 percent but not the 95 confidence level. While these results are slightly more constraining than what we obtain with the {\tt KiDS} data, they show that current weak-lensing surveys are not yet capable of clearly distinguishing between different feedback models.

The situation is, however, very different when weak-lensing observations are combined with data from the X-ray gas fractions and kSZ profiles. In this case we obtain much more constraining results that strongly disagree with all simulations featuring mild baryonic feedback effects (such as {\tt EAGLE}, {IllustrisTNG}, {\tt H-AGN}, and {\tt MassiveBlackII}). Note that similar (yet somewhat less constraining) results have been reported in \citet{Troster:2021gsz}.

Finally, we would like to note that the results illustrated in Fig.~\ref{fig:ps_constraints} are not only important in the context of weak lensing but have consequences for other cosmological studies. They show that any cosmological analysis pipeline that relies on the matter power spectrum above  $k\sim0.1$ h/Mpc will have to account for baryonic feedback processes. This includes for example spectroscopic and photometric galaxy clustering or 21-cm intensity mapping observations.

\section{Conclusions}\label{sec:conclusions}
In this paper we carry out an independent analysis of the {\tt KiDS}-1000 weak lensing (WL) band power measurements and we combine, for the first time, WL data with observations from X-ray gas (and stellar) fractions and with stacked gas profiles from the kSZ effect. We thereby specifically focus on the influence of baryonic feedback processes, which are modelled using the baryonification method and assuming seven free parameters.

There are three main questions we attempt to answer with the present paper: (i) how well can we constrain the parameters of the baryonification model with current data? (ii) what does the data tell us about the baryonic suppression of the matter power spectrum? (iii) Does a more general prescription for the baryonic effects alter current constraints on cosmological parameters and can it alleviate the observed $S_8$-tension? We are now summarising the answers to these three main questions:
\begin{itemize}
    \item[(i)] The baryonification model provides a more versatile and generally valid description of baryonic effects compared to what has been used before (such as in the original {\tt KiDS} analysis). The model contains seven free parameters that can be directly constrained with observations, and it is able to reproduce results from a plethora of different hydrodynamical simulations.  We show that the data from the WL band-power alone is unable to constrain the baryonification parameters beyond their prior ranges. Combining weak lensing with X-ray and kSZ observations, on the other hand, allows us to put constraints on three of the seven baryonic parameters ($M_c$, $\mu$, $\eta$). For two more parameters, we obtain meaningful trends ($\gamma$, $\delta$), while the last two remain largely unconstrained ($\theta_{\rm ej}$, $\eta_{\delta}$).
    \item[(ii)] Quantifying the baryon-induced suppression of the matter power spectrum ($\mathcal{S}$) is of interest not only for weak lensing but also for other cosmological observations. Furthermore, it can be used as a discriminant of hydrodynamical simulations and their feedback implementation schemes. Assuming the full WL + Xray + kSZ data set and restricting the cosmic baryon fraction to $f_b=0.15-0.17$ (a range that includes both {\tt Planck} and {\tt WMAP} cosmologies), we find the minimum of the baryonic suppression signal to be at $\mathcal{S}_{\rm min}=0.67-0.79$, which corresponds to a 21-33 percent baryonic suppression effect (at 95 percent confidence level). For the largest scale (smallest wave mode $k_{\rm min}$) where $\mathcal{S}$ deviates from unity by more than one percent, we find $k_{\rm min}=0.1-0.45$ h/Mpc (at 95 percent confidence level). This corresponds to a significantly larger scale than what is predicted by the majority of hydrodynamical simulations.
    \item[(iii)] Baryonic feedback processes affect the clustering strength and therefore the observed tension between the {\tt KiDS}-1000 WL band power measurements and the CMB data from {\tt Planck} (often referred to as the $S_8$-tension). We report the discrepancy between the two data sets to decrease from 3.8 to 2.9 $\sigma$ when baryonic suppression effects are included into the analysis. Furthermore, we find the external X-ray and kSZ data sets to favour strong baryonic feedback. The combined WL + Xray + kSZ data vector yields a remaining tension of 2.6 $\sigma$ with the data from {\tt Planck}. We conclude that strong baryonic feedback effects may help to alleviate the $S_8$ tension without, however, fully solving it.
\end{itemize}
The results from this paper further highlight the importance of including the effects of baryons into the analysis pipelines of modern WL surveys. In the future, we foresee that a combination of WL observations with data from X-ray and the SZ effect will allow us to account for baryonic nuisance parameters without biasing or significantly increasing the errors of cosmological parameters.

\section*{Acknowledgements}
We thank Tilman Troester for his help with the analysis of the {\tt KiDS}-1000 data. This work is supported by the Swiss National Science Foundation via the grant {PCEFP2\_181157}.

%%%%%%%%%%%%%%%%%%%%%%%%%%%%%%%%%%%%%%%%%%%%%%%%%%
\section*{Data Availability}
The data underlying this article will be shared on reasonable request to the corresponding author.
 
%The inclusion of a Data Availability Statement is a requirement for articles published in MNRAS. Data Availability Statements provide a standardised format for readers to understand the availability of data underlying the research results described in the article. The statement may refer to original data generated in the course of the study or to third-party data analysed in the article. The statement should describe and provide means of access, where possible, by linking to the data or providing the required accession numbers for the relevant databases or DOIs.



\bibliographystyle{mnras}
\bibliography{references} % if your bibtex file is called example.bib

%\begin{thebibliography}{99}
%\end{thebibliography}

%%%%%%%%%%%%%%%%%%%%%%%%%%%%%%%%%%%%%%%%%%%%%%%%%%

%%%%%%%%%%%%%%%%% APPENDICES %%%%%%%%%%%%%%%%%%%%%

\appendix

\section{Comparison with the original {\tt KiDS} analysis}\label{app:comparisonKiDS}
Throughout this paper, we use the weak-lensing band power data vector and covariance matrix provided by the {\tt KiDS}-1000 science team \citep{KiDS:2020suj}. However, our analysis pipeline differs from the one used in the original {\tt KiDS} analysis (see discussion in Sec.~\ref{sec:data}). Most notably, in order to predict the matter power spectrum, we rely on the revised halofit method \citep{Takahashi:2012em} instead of the HMcode \citep{Mead:2015yca}. Furthermore, our initial transfer function is obtained using the \cite{Eisenstein:1997ik} fitting function (instead of the Boltzmann solver {\tt Class}) and we neglect any effects from massive neutrinos.

Next to the differences in the prediction pipeline, we do not vary the same parameters during our MCMC sampling runs. Most notably, we use the baryonification model with seven free baryonic parameters as opposed to one parameter used by the {\tt KiDS} team. On the other hand, we abstain from varying any parameters related to the redshift distribution of the source galaxies.

\begin{figure} 
\centering
\includegraphics[width=0.48\textwidth,trim=0.2cm 0.1cm 1.0cm 1.0cm, clip]{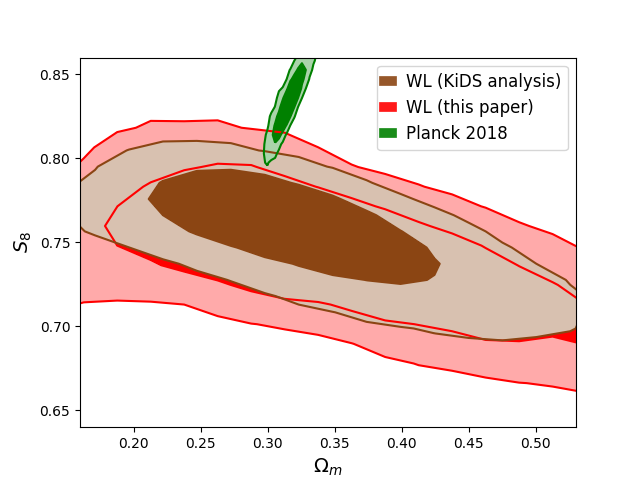}
\caption{Comparison of the posterior distribution from our analysis (red) and the {\tt KiDS} results (brown). The contours from {\tt Planck} are shown in green for reference. The details about the difference between the two analysis pipelines are given in the text.} 
\label{fig:OmS8comp}
\end{figure}

In Fig.~\ref{fig:OmS8comp} we compare the $S_8-\Omega_m$ posteriors from this paper (red) with the ones obtained by the {\tt KiDS} team (brown). Our contours are broader, extending further towards small and larger values of $S_8$. We speculate this to be an effect of the increased number of baryonic parameters, which also allows for scenarios featuring a baryonic boost (instead of a suppression) of the power spectrum. Compared to the posterior from {\tt KiDS}, our contours are also slightly shifted downwards. The effect is most visible when focusing on the 68 percent confidence region. This shift could be a consequence of using the revised halofit model instead of HMcode. Indeed, it has been shown using {\tt KiDS} 450 data that moving from HMcode to halofit may lead to a bias on $S_8$ of up to 0.3 standard deviations \citep{Joachimi:2020abi}.

\section{Baryonic parameter constraints from specific data combinations}\label{app:bfcconstraints}
\begin{figure*} 
\centering
\includegraphics[width=0.95\textwidth,trim=0.2cm 0.1cm 0.9cm 0.5cm, clip]{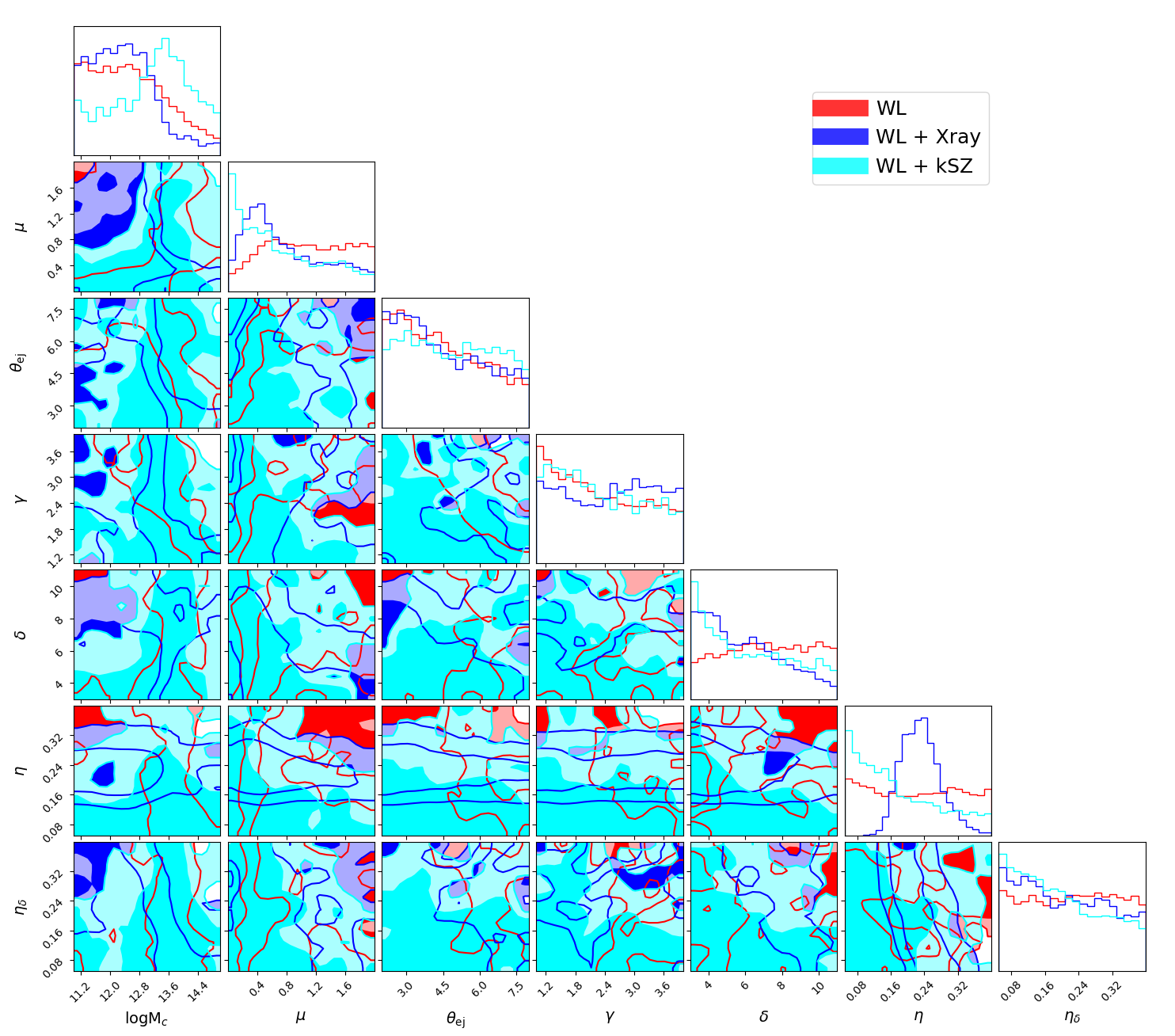}
\caption{Posterior contours of the baryonification model parameters at 68 and 95 percent confidence level. Results based on {\tt KiDS}-1000  weak-lensing data alone are shown in red. Results from the combined WL + Xray and WL + kSZ data sets are are shown in blue and cyan, respectively. Posterior contours from the full WL + Xray + kSZ data combination are illustrated in Fig.~\ref{fig:bfc_constraints} of the main text.}
\label{fig:bfc_constraints2}
\end{figure*}

In Sec.~\ref{sec:results} we show the posterior contours of the seven baryonic parameters using data from the {\tt KiDS}-1000 WL band power data alone and from the combined WL + Xray + kSZ data vector. In this appendix, we also discuss the intermediate cases, namely the WL + Xray and the WL + kSZ data combinations. This allows us to investigate more subtle differences between the results from the X-ray gas (and stellar) fractions, on the one hand, and the stacked kSZ profiles, on the other hand.

Fig.~\ref{fig:bfc_constraints2} shows the posterior contours of the baryonification parameters for the WL only (red), the WL + Xray (blue), and the WL + kSZ (cyan) data sets. The WL only case has already been shown in Fig.~\ref{fig:bfc_constraints} and is added here as a reference. 

There are several notable differences between the blue and the cyan posteriors. Most importantly, the $M_c$-parameter, describing the mass-dependence of the gas profile, is centred around $\sim 10^{12}$ M$_{\odot}$/h in the WL + Xray and around $\sim4\times 10^{13}$ M$_{\odot}$/h in the WL + kSZ case.
Another slight difference can be observed in the behaviour of $\mu$, the parameter describing how strongly low-mass and high-mass haloes differ in their gas profile. While in the WL + Xray case, very small values of $\mu$ are disfavoured, they are preferred in the WL + kSZ scenario.
Finally, there is a visible difference between the blue and cyan contours of the stellar parameter $\eta$, which is well constrained in the WL + X-ray and completely unconstrained in the WL + kSZ case. This is a direct consequence of the stellar fractions that are part of the X-ray data set. All other parameters shown in Fig.~\ref{fig:bfc_constraints2} behave similarly in both cases.

%%%%%%%%%

% Don't change these lines
\bsp	% typesetting comment
\label{lastpage}
\end{document}